\documentclass[12pt]{article}
\newcommand{\sect}[1]{\setcounter{equation}{0}\section{#1}}

\newcommand{\be}{\begin{equation}}
\newcommand{\ee}{\end{equation}}
\newcommand{\bea}{\begin{eqnarray}}
\newcommand{\eea}{\end{eqnarray}}
\newcommand{\p}{\partial}

\newcommand{\e}{\epsilon}

\begin{document}
\renewcommand{\thefootnote}{\fnsymbol{footnote}}
\begin{titlepage}
\begin{flushright}
%ROM2F/2004/19\\
1207.5459 [hep-th]
\end{flushright}
\vskip .7in
\begin{center}
{\Large \bf On the D-brane solutions in G$\rm {\ddot o}$del
Universe} \vskip .7in {\large Kamal L.
Panigrahi{$^{a,b}$}}\footnote{e-mail: {\tt panigrahi@iitrpr.ac.in,
panigrahi@phy.iitkgp.ernet.in}}, {\large Pratap K.
Swain{$^b$}}\footnote{e-mail:{\tt pratap@phy.iitkgp.ernet.in}}
\vskip .2in {$^a$} {\it Department of Physics,\\
Indian Institute of Technology Ropar, Rupnagar-140001, INDIA}\\
\vskip .1in
{$^b$}{\it Department of Physics and Meteorology, \\
Indian Institute of Technology Kharagpur,
Kharagpur-721302, INDIA}\\
\vspace{.7in}
\begin{abstract}
\vskip .5in
\noindent
We present a class of supersymmetric G$\rm {\ddot o}$del solutions in string theory from the non-standard
intersection of branes in supergravities. Such solutions are obtained by applying a T-duality
on the known solutions in PP-wave spacetime. We further present classical solutions of
supersymmetric D-brane in G$\rm {\ddot o}$del universes arising from the PP-wave in the near horizon geometry of
stack of D5-branes and from the new isometries of $H_{6}$ PP-wave background.
These branes are supported by multiple constant Neveu-Schwarz and Ramond-Ramond field strengths.
\end{abstract}
\end{center}
\vfill
\end{titlepage}
\setcounter{footnote}{0}
\renewcommand{\thefootnote}{\arabic{footnote}}
\sect{Introduction and summary} G$\rm {\ddot o}$del universe is a
homogeneous rotating cosmological solution of Einstein's equations
with pressureless matter and negative cosmological constant, which
played an important role in the conceptual development of general
theory relativity. In \cite{Gauntlett:2002nw} an M-theory solution
of G$\rm {\ddot o}$del universe type has been found out and it was
shown to preserve 20 supersymmetries. Furthermore it generates
Ramond-Ramond (RR) fluxes when compactified down to 10-dimensional
type IIA string theory. They contain some unphysical features like
the closed time like curves (CTC), but the problem was resolved
geometrically in \cite{Herdeiro:2002ft} in the context of spinning
deformation of (D1-D5) system. Further in \cite{Boyda:2002ba}, it
was argued that the principle of holography remedy this problem
and protect the chronology in the G$\rm {\ddot o}$del universe
background. It was further shown that they are related to PP-wave
background by a $T$-duality transformation. In this connection in
\cite{Harmark:2003ud} a large class of solutions were found out in
the context of string theory in PP-wave background and
corresponding properties of such spacetime including that of
supersymmetries has been analyzed in details both in 10 and 11
dimensions. These class of solutions were obtained by applying $T$
and $S$-duality transformations in the relevant solutions in
PP-wave background. The string theory spectrum has also been
studied by invoking the idea of quantization of PP-wave in light
cone gauge. Further it was also noticed that the supergravity
solutions of D5-branes in a type IIB PP-wave
background\cite{Kumar:2002ps} (coming from $AdS_3\times S^3$
geometries) after a T-duality transformation can give new
localized D4-brane solutions in the G$\rm {\ddot o}$del universe.
They were also obtained by looking at the relevant boundary
conditions in the open string constructions in
PP-wave\cite{Skenderis:2002wx} and then applying T-dualities. This
generates the mixed boundary conditions. Also in
\cite{Brace:2003st}, using the duality described in
\cite{Boyda:2002ba}, the string quantization has been studied in
the ten dimensional description of these solutions and yet another
mechanism has been proposed to resolve the CTCs within string
theory. In this paper we would like to obtain new supersymmetric
G$\rm {\ddot o}$del backgrounds by applying T-dualities in a class
of PP-wave background which was derived from the Penrose limit on
the non-standard intersection of D-branes in supergravities.
Indeed in \cite{Lu:2002kw} a large class of PP-wave backgrounds
were obtained from the non-standard brane intersections whose near
horizon geometry was essentially Anti-de Siter (AdS). The
resulting PP-waves are shown to be supported by multiple constant
RR and NS-NS field strengths and they are interesting in their own
right. These solutions are different from the known  near horizon
and PP-wave limit of the usual intersecting (like D1-D5) branes in
supergravities. The main difference stems from the fact that in
this case the Harmonic functions of branes depend on the relative
transverse space. We have found out the new G$\rm {\ddot o}$del
universe backgrounds by applying T-dualities on these class of
PP-wave solutions. We take two examples, the (D1/D5/D5) system and
the D3/D5/D5/ND5/NS5 intersecting brane system and have obtained
new G$\rm {\ddot o}$del universe backgrounds. These solutions are
different from the known solutions as they contain multiple RR and
NS-NS fields, but keeps the ``G$\rm {\ddot o}$del structure'' is
still intact. Next, we find out some new supergravity solutions of
D-branes in type IIA string theory from the known solutions of
D-branes in PP-wave backgrounds\footnote{D-brane supergravity
solutions in NS-NS and R-R PP-waves have been discussed in, for
example in
\cite{Kumar:2002ps}\cite{Bain:2002nq}\cite{Alishahiha:2002rw}\cite{Biswas:2002yz}\cite{Nayak:2002ty}
\cite{Panigrahi:2003rh}\cite{Ohta:2003rr}\cite{Hassan:2003ec}}.
First we have taken the example of D5-branes in the near horizon
and PP-wave backgorund of a stack of D5-branes in type IIB string
theory. This solution was found in \cite{Biswas:2002yz} and was
shown to be 1/4 supersymmetric. We apply T-duality and present a
D4-brane solution in G$\rm {\ddot o}$del universe (the so-called
$n=1$ G${\rm \ddot o}$del model) and examine the fate of unbroken
supersymmetry by solving the gravitino and dilatino variations
explicitly. Further we have uplifted this solution to M5-brane in
M-theory. Our next example is a D2-brane in G$\rm {\ddot o}$del
universe which is obtained from a D3-brane in PP-wave in the
presence of both RR and NS-NS 3 form field strengths found in
\cite{Alishahiha:2002rw}. The rest of the paper is organized as
follows. In section-2, we find out new supersymmetric G$\rm {\ddot
o}$del solutions from intersecting branes whose near horizon
geometry are of AdS type. In section-3, we find out new
supergravity solutions for D-branes in G$\rm {\ddot o}$del
universes from the corresponding branes in PP-wave backgrounds.
Section-4 is devoted to the analysis of unbroken spacetime
supersymmetry. Finally in section-5, we conclude with some
remarks.

 \sect{G$\rm {\ddot o}$del Universes from intersecting branes} In
this section we will find out new G$\rm {\ddot o}$del models from
the PP-wave background of non-standard brane intersections in
supergravities. The AdS structure in the near horizon geometry of
such intersections arises from the fact the Harmonic function for
each participating brane depends on the relative transverse space
rather than the overall transverse space. The first example we
consider is the intersecting (D1/D5/D5)-brane system that couple
to three form R-R field strengths in D=10. The relevant metric and
other fields are given by \cite{Cowdall:1998bu} \bea
ds^2_{10} &=& G^{-3/4}(F\tilde F)^{-1/4} \left(-dt^2 + dx^2 + G F dy^2_i + G\tilde F d{\tilde y}^2_i \right)
\nonumber \\
F_{(3)} &=& e^{\phi} * (\tilde F dt \wedge dx \wedge d^4{\tilde y} \wedge dF^{-1})
+ e^{\phi} * (F dt \wedge dx \wedge d^4 y \wedge d{\tilde F}^{-1}) \nonumber \\
&+& dt\wedge dx\wedge dG^{-1} \ , \>\>\> e^{2\phi} = \frac{F\tilde
F}{G} \ , \>\>\> G = F\tilde F \ , \label{d1-d5-d5} \eea where
$y_i$ and ${\tilde y}_i$ are the coordinates in the relative
transverse space of the stack of D5-branes, $F = 1 +
\frac{Q}{y^2}$, $\tilde F = 1 + \frac{Q}{\tilde y^2}$ are the
harmonic functions of the branes. This particular solution has the
property that the dilaton vanishes. The near horizon geometry of
such a solution is $AdS_3 \times S^3 \times S^3 \times S^1$. The
PP-wave geometry was obtained in \cite{Lu:2002kw}, after taking a
suitable Penrose limit on (\ref{d1-d5-d5}) and it is given by \bea
ds^2_{10} &=& - 2 dx^+ dx^- + H {(dx^+)}^2 + \sum^8_{i =1} dx^2_i \ , \>\>\> F^{(RR)}_{3} = dx^+ \wedge \Phi_{(2)} \ , \nonumber \\
H &=& -\mu^2 (x^2_1 + x^2_2) - \frac{\mu^2}{2} \cos^2 \alpha (x^2_3 + x^2_4) - \frac{\mu^2}{2} \sin^2 \alpha (x^2_5 + x^2_6) \ , \nonumber \\
\Phi_{(2)} &=& 2 \mu dx_1 \wedge dx_2 + \sqrt{2}\mu\cos\alpha ~ dx_3 \wedge dx_4 - \sqrt{2}\mu \sin\alpha ~dx_5 \wedge dx_6 \ ,
\eea
where $\alpha$ is the angle of rotation between the coordinates of two spheres in the
transverse space of branes.
We wish to find out the G$\rm {\ddot o}$del spacetime of this geometry. We shall follow the
same procudure of getting a G${\rm{\ddot o}}$del from PP-wave by applying a T-duality
as described in \cite{Harmark:2003ud}. The first step is to do the following coordinate transformation
\begin{eqnarray}
x^+ = x^0 + x^9, ~ x^- =\frac{x^0-x^9}{2}, \> x^{2k-1} + ix^{2k}
\rightarrow (x^{2k-1} + i x^{2k}) e^{-i\mu_k x^+}, \> k = 1,2,3.
\end{eqnarray}
With the above transformation, the new metric looks like
\begin{eqnarray}
ds^2 = - {(dx^0)}^2 + {(dx^9})^2  + \sum_{i=1}^8dx_i^2 -2\sum_{i,j=1}^6 J_{ij}x_i dx_j(dx^0 + dx^9) \ ,
\end{eqnarray}
where
\begin{eqnarray}
J_{12} = \mu = - J_{21} \ , \>\>
J_{34} = \frac{\mu}{\sqrt{2}}\cos\alpha = - J_{43} \ , \>\>
J_{56} = \frac{\mu}{\sqrt{2}}\sin\alpha = - J_{65} \ .
\end{eqnarray}
The next step is to apply a $T$-duality along
the $x^9$ direction \footnote{The T-duality transformation can be found out for example in \cite{Breckenridge:1996tt}}.
The new metric and fields after the $T$-duality become
\begin{eqnarray}
ds^2 &=& - (dx^0 + \sum_{i,j=1}^6 J_{ij}x_i dx_j)^2 + (dx^9)^2 + \sum_{i=1}^9dx_i^2 \ ,
H_{129} = -F_{0129} = F_{12}= -2\mu \ , \nonumber \\
H_{349} &=& -F_{0349}= F_{34}= -\sqrt{2}\mu\cos\alpha ,\>\>  ~ F_{0569} =  H_{569} = - F_{56}
= -\sqrt{2}\mu\sin\alpha .
\end{eqnarray}
This background is different from the known examples of
\cite{Harmark:2003ud}. It is also important to note that this
background preserves 1/2 supersymmetry only for $\alpha = \pi/4$
which is expected from \cite{Lu:2002kw}. Our next example is a
non-standard intersection of D3/D5/D5/NS5/NS5 branes. This near
horizon geometry was found out to be $AdS_3 \times S^2 \times S^2
\times T^3$. The Penrose limit was taken and the resulting pp-wave
background has been written in \cite{Lu:2002kw}. We wish to write
it once for our future reference
 \bea ds^2 &=& -2 dx^+ dx^- -
\mu^2 \left(x^2_2 + x^2_2 + 2 \cos^2 \alpha x^2_3 + 2 \sin^2
\alpha x^2_4\right) {(dx^+)}^2
+ dx^2_i \ , \nonumber \\
F_{+1268} &=&  2\mu  \ , \>\>\> F_{+36} = H_{+38} = \sqrt{2}\mu
\cos \alpha \ , \>\>\> F_{+48} = H_{+46} = -\sqrt{2}\mu \sin\alpha
\ . \eea For our purpose we will set $\alpha = \pi/4$. With this
choice, the metric and other fields can be read off as \bea ds^2
&=& -2dx^+dx^- -\mu^2\sum^{4}_{i=1} x^2_i {(dx^+)}^2 +
\sum_{i=1}^8
{dx_i}^2 \ , \nonumber \\
F_{+1268} &=&  2\mu  \ , \>\>\> F_{+36} = H_{+38} = \mu \ , \>\>\>
F_{+48} = H_{+46} = - \mu  \ . \eea After applying $T$-duality
along $x^9$ direction as described earlier, we end of with the
following form of the metric and other resultant field strengths
as
\begin{eqnarray}
ds^2 &=& - (dx^0 + \sum_{i,j=1}^4 J_{ij}x_i dx_j)^2 + (dx^9)^2  +
\sum_{i=1}^8dx_i^2 \ , ~~ H_{129} = H_{349} = -2\mu, \cr & \cr
F_{0369} &=& \mu = - F_{0489} \ , ~~ F_{1268} = F_{3457} = 2\mu \
,~~ F_{36} = -\mu = - F_{48} \ , \cr & \cr H_{038} &=& \mu =
H_{938} \ , ~~ H_{046} = -\mu = H_{946}
\end{eqnarray}
\sect{D-brane solution in G$\rm {\ddot o}$del Universes} In this
section, we would like to write down the D4-brane solutions in the
G${\rm{\ddot o}}$del universe of $n=2$ type presented in
\cite{Harmark:2003ud}, which will be used in the next section to
study the supersymmetry. The metric, dilaton and various field
strengths of a stack of D4-branes is given by
\cite{Harmark:2003ud}
\begin{eqnarray}
ds^2 &=& f^{-1/2}_4\left(-(dt + \mu \sum^4_{i =1} J_{ij} x^i
dx^j)^2  + \sum^4_{i =1} {(dx^i)}^2\right)
+ f^{1/2}_4 \sum^9_{m=5} {(dx^m)}^2 \nonumber \\
e^{2\phi} &=& f^{-1/2}_4 \ , \>\>\>\> F_{12} = F_{34} = - 2\mu \ ,
\>\>\> H_{129} = H_{349} = -2\mu \ , \>\>\>
F_{0129} = F_{0349} = 2\mu \ , \nonumber \\
F_{mnpq} &=& \epsilon_{mnpqr}\p_r f_4 \ ,\>\>\> f_4 = 1 + \frac{N
g_s l^3_s}{r^3} \ , \>\>\> r^2 = \sum^9_{m=5}{(x^m)^2} ,
\label{godel-D4-n2}
\end{eqnarray}
where $J_{12} = -J_{21} = J_{34} = -J_{43} = 1$ and $f_4$ is the
harmonic function of the D4-brane in the transverse five-space.
One can observe that the presence of various field strengths
symmetrically along the $x^1, x^2$ and correspondingly along the
$x^3, x^4$ directions. We will see that this structure plays a
crucial role in the supersymmetry analysis of the D4-brane
solution. Now we would like to present further examples of D-brane
solution in the G$\rm {\ddot o}$del universe models. Our first
example is a D4-brane in the so-called $n=1$ G$\rm {\ddot o}$del
universe model. This is obtained by applying $T$-duality along
isometry directions of the D5-brane in a PP-wave background that
arises from the near horizon and Penrose limit of a stack of
coincident D5-branes and is dual to the PP-wave background of
Nappi-Witten model. The supergravity solution of D-branes were
presented in \cite{Biswas:2002yz}. In particular the 1/4
supersymmetric D5-brane is written as \cite{Biswas:2002yz} \bea
ds^2 &=& f^{-1/2}_5\left(-2dx^+ dx^- - \mu^2\sum^{2}_{i=1} x^2_i
{(dx^+)}^2 + \sum^4_{a =1} {(dx^a)}^2\right)
+ f^{1/2}_5 \sum^8_{m=5} {(dx^m)}^2 , \nonumber \\
e^{2\phi} &=& f_5^{-1}, \>\>\>\> F_{+12} = 2\mu, \>\>\>\> F_{mnp}
= \epsilon_{mnpq}\p_q f_5,\>\>\> f_5 = 1 + \frac{Ng_s l^2_2}{r^2},
\>\> r =\sqrt{\sum^8_{m=5} {(x^m)}^2} \ . \nonumber
\\ \eea
The spacetime supersymmetry was analyzed by solving the dilatino
and gravitino variations explicitly and it was found out that in
addition to the flat space D5-brane supersymmetry condition if a
`necessary' condition $\Gamma^{\hat +} \epsilon = 0$ acts on the
killing spinors, then all variations are satisfied giving a
solution for the spinors which preserves eight supercharges.
%We
%wish to apply a T-duality along the direction $x^9$, where $x^{+}
%= x^0 + x^9$, and $x^- = (x^0 - x^9)/2$ to get a D4-brane solution
%in G$\rm {\ddot o}$del model.
Applying a $T$-duality along $x^9$ as described in the last
section, we get the following form of the metric, field strengths
and dilaton for the `localized' D4-brane in G$\rm {\ddot o}$del
model.
 \bea
ds^2 &=& f^{-1/2}_4\left(-(dt + \mu \sum^2_{i =1} J_{ij} x^i
dx^j)^2  + \sum^4_{a =1} {(dx^a)}^2\right)
+ f^{1/2}_4 \sum^9_{m=5} {(dx^m)}^2 \nonumber \\
e^{2\phi} &=& f^{-1/2}_4 \ , \>\>\>\> F_{12} = - 2\mu \ , \>\>\> H_{129} = -2\mu \ , \>\>\>
F_{0129} = 2\mu \ , \>\>\>\> F_{mnpq} = \epsilon_{mnpqr}\p_r f_4 \ , \nonumber \\
J_{12} &=& 1 = -J_{21} \ , \>\>\> f_4 = 1 + \frac{N g_s
l^3_s}{r^3} \ , \>\>\> r^2 = \sum^9_{m=5}{(x^m)^2} .
\label{godel-D4} \eea We have checked that the solution presented
above solves all type-IIA field equations. Next we would like to
get a M5-brane solution starting from the D4-brane solution
presented above in the G$\rm {\ddot o}$del model. Using the well
known relation between the 10d and 11d metric: \bea ds^2_{11} =
e^{-\frac{2\Phi}{3}} ds^2_{10} + e^{\frac{4\Phi}{3}} (dx_{11} +
A_{\mu} dx^{\mu})^2, \eea where $ds^2_{11}$ and $ds^2_{10}$
represent the metric in eleven and ten dimensions respectively,
and $A_{\mu}$ is the one-form field (which is zero in the present
case). One can easily see that the M5-brane solution is given by
\bea ds^2 &=& f^{-1/3}\left(-2 dx^+ dx^- - \mu^2 \sum^{2}_{i =1}
{(x^i)}^2 {(dx^+)}^2
+ \sum^4_{a =1}{(dx^a)}^2 \right) + f^{2/3}\sum^9_{m =5}{(dx^m)}^2  \ , \nonumber \\
F_{+129} &=& 2\mu \ , \>\>\> F_{mnpq} = \epsilon_{mnpqr}\p_r f \ ,
f = 1+ \frac{Nl^3_p}{r^3}, \eea with $l_p$ being the eleven
dimensional Plank length. In writing down the above solution in
the $x^+, x^-$-coordinates, we have made the following change of
variables
\bea x^1 + i x^2 \rightarrow (x^1 + i x^2)e^{-2\mu x^+}
\ . \eea
The solution can directly be obtained from the PP-wave
solution by uplifting it to eleven dimensions. Note that in
absence of any D-brane charges, if we apply $T$ dualities along
$x^3$ and $x^4$ directions, we get the follwing form of metric and
RR fields
\bea ds^2 &=& -2dx^+ dx^- - \mu^2 \sum^2_{i=1} {(x^i)}^2
{(dx^+)}^2 + \sum^8_{m=1} {(dx^m)}^2 \ , \nonumber \\ F_{+1234}
&=& F_{+5678} = 2\mu . \eea
 Once again by applying a $T$-duality
along the $x^9$ direction as before we get the following Godel metric and
other field strengths
\bea ds^2 &=& -\left[dt + \mu(x^1 dx^2 - x^2
dx^1)\right]^2 + \sum^9_{m=1} {(dx^m)}^2 \, \nonumber \\ F_{1234}
&=& F_{5678} = 2\mu, \>\>\> H_{129} = 2\mu \ , \eea
Next we would
like to find a D2-brane solution in a G$\rm {\ddot o}$del model.
The D2-brane can be obtained by applying a T-duality along a
localized D3-brane solution described in \cite{Alishahiha:2002rw}.
Note that this D3-brane solution was obtained by applying
succssive $T$-dualities along the new isometry directions of the
localized D5-brane solution of \cite{Alishahiha:2002rw} by following
\cite{Michelson:2002wa}. In stead of going into the detials of
construction we present here the final form of D3-brane solution in the presence of
various R-R and NS-NS fluxes as
\bea ds^2 &=& f^{-1/2}_3
\left(-2dx^+ dx^- - 4\mu^2 [x^2_1 + x^4_2] {(dx^+)}^2 + dx^2_1 +
dx^2_2 \right)
+ f^{1/2}_3 \left(dr^2 + r^2 d\Omega^2_5\right) \ , \nonumber \\
F_{+31} &=& F_{+42} = 2\mu \ , \>\>\> H_{+41} = H_{+32} = 4\mu   \
, \>\>\> F_{mnpqr} = \epsilon_{mnpqrs}\p_{s} f_3 \ , \>\>\> f_3 =
1 + \frac{N g_s l^4_s}{r^4} \ ,
\nonumber \\
\eea where $f_3$ is harmonic function in the transverse six space.
By applying $T$-duality along $x^9$-direction as before, we get
the following metric and other field strengths \bea ds^2 &=&
f^{-1/2}_2\left[-{\left(dx^0 + 2\mu \sum^{2}_{i,j=1} J_{ij} x^i
dx^j\right)}^2 + \sum^{2}_{i=1} {(dx^i)}^2\right]
+ f^{1/2}_2\sum^{9}_{m = 3} {(dx^m)}^2\ , \nonumber \\
e^{2\phi} &=& f^{1/2}_2 \ , \>\>\> A_{012} = f^{-1}_2 \ , \>\>\> F_{0329} = F_{0429} = 2\mu \ , F_{31} = F_{42} = -2\mu \ , \nonumber \\
H_{041} &=& H_{941} = 4\mu = - H_{129} \ , H_{032} = H_{932} = 4\mu \ , f_2 = 1 + \frac{Ng_s l^5}{r^5} \ ,
r^2= \sum^9_{m=3} {(x^m)}^2 \ . \nonumber \\
\label{godel-D2} \eea Once again we have checked that the
localized D2-brane solution above solves all type-IIA field
equations of motion. Other D-branes and their bound states can be
found out by applying $T$-dualities along various isometries of
the solution presented here. \sect{Spacetime Supersymmetry
Analysis} In this section, we will analyze the the fate of the
unbroken spacetime supersymmetry of the D-brane solutions
presented above by solving the dilatino and gravitino variations
explicitly. The supersymmetry variation of the dilatino and
gravitino fields in type IIA supergravity in string frame is given
by \cite{Schwarz:1983qr,Hassan:1999bv}.
\begin{eqnarray}
\delta \lambda = {1\over2}(\Gamma^{\mu}\partial_{\mu}\Phi
- {1\over 12} \Gamma^{\mu \nu \rho}H_{\mu \nu
\rho})\epsilon + {1\over
  8}e^{\Phi}(5F^{(0)} - {3\over 2!} \Gamma^{\mu \nu}F^{(2)}_{\mu \nu } + {1\over 4!} \Gamma^{\mu \nu \rho \sigma} F_{\mu \nu \rho \sigma})\epsilon,
\label{dilatino}
\end{eqnarray}
\begin{eqnarray}
\delta {\Psi_{\mu}} = \Big[\partial_{\mu} + {1\over
8}(w_{\mu
  \hat a \hat b} -  H_{\mu \hat{a}
  \hat{b}})\Gamma^{\hat{a}\hat{b}}\Big]\epsilon
+ {1\over 8}e^{\Phi}\Big[F^{(0)} - {1\over 2!} \Gamma^{
\mu \nu}F^{(2)}_{\mu \nu} + {1\over 4!}
\Gamma^{\mu \nu \rho \sigma}F^{(4)}_{\mu \nu \rho
\sigma}\Big]\Gamma_{\mu}\epsilon, \nonumber \\
\label{gravitino}
\end{eqnarray}
where we have used $\mu, \nu, \rho$ to describe the ten
dimensional space-time indices, and the hated ones are the
corresponding tangent space indices. Solving the Dilatino
variation (\ref{dilatino}) for the D4-brane solution
(\ref{godel-D4-n2}), presented in \cite{Harmark:2003ud} we get the
following condition on the spinors to be satisfied \bea &&f^{-5/4}
f_{,m} \left(\Gamma^{\hat m} + \frac{1}{4!}\epsilon_{\hat m\hat
n\hat p\hat  q\hat r} \Gamma^{\hat n \hat p \hat q \hat
r}\right)~\epsilon -2 \mu f^{1/4} \left(\Gamma^{\hat 1 \hat 2} +
\Gamma^{\hat 3 \hat 4}\right)\Gamma^{\hat 9}\epsilon \cr & \cr &&
- \mu f^{1/4} \left(\Gamma^{\hat 1 \hat 2} + \Gamma^{\hat 3 \hat
4}\right)\epsilon - \mu f^{-1/4} \Gamma^{\hat 0}\left(\Gamma^{\hat
1 \hat 2} + \Gamma^{\hat 3 \hat 4}\right)\Gamma^{\hat 9} \epsilon
= 0 \, \eea Now solving the gravitino variations we get the
following \bea \p_0 \e = 0, \>\>\> \p_a \e = 0,\>\>\> (a=5,\cdots,
9), \>\>\> \p_i \e + \frac{\mu}{2}J_{ij}\Gamma^{\hat j \hat 9} \e
= 0, \>\>\> (i =1,\cdots, 4) \ . \label{gravi-d4}\eea Note that
while writing down the above variations (\ref{gravi-d4}) we have
made use of the D4-brane supersymmetry condition in flat space
\bea \left(\Gamma^{\hat m} + \frac{1}{4!}\epsilon_{\hat m\hat
n\hat p\hat q\hat r} \Gamma^{\hat n \hat p \hat q \hat
r}\right)~\epsilon = 0 \ , \label{D4-susy} \eea and \bea \left(1-
\Gamma^{\hat 1 \hat 2 \hat 3 \hat 4}\right) ~\epsilon = 0 \ . \eea
By using the above two conditions all the dilatino and gravitino
variations are satisfied leaving only 1/4 of the total spacetime
supersymmetry unbroken and is solved by a constant spinor. Hence
the D4-brane solution in the $n=2$ G$\rm{\ddot o}$del universe
model preserves 1/4 unbroken supersymmetry. Let us now look at the
fate of the unbroken supersymmetry for the D4-brane in $n=1$
G$\rm{\ddot o}$del model. First, solving the dilatino variation
(\ref{dilatino}), for the D4-brane solution presented in
(\ref{godel-D4}) we get  \bea &&f^{-5/4} f_{,m} \left(\Gamma^{\hat
m} + \frac{1}{4!}\epsilon_{mnpqr} \Gamma^{\hat m \hat n \hat p
\hat q}\right)~\epsilon -2\mu f^{1/4} \Gamma^{\hat 1 \hat
2}\left(\Gamma^{\hat 9} + \frac{1}{2}(1 + \Gamma^{\hat 0 \hat
9})\right)\epsilon = 0 \, \label{ggd4} \eea The vanishing of the
dilatino variation demands that the following two conditions to be
imposed \bea \left(\Gamma^{\hat m} + \frac{1}{4!}\epsilon_{\hat
m\hat n\hat p\hat q\hat r} \Gamma^{\hat n \hat p \hat q \hat
r}\right)~\epsilon = 0 \label{flat-D4} \eea and \bea \Gamma^{\hat
0} \e = \Gamma^{\hat 9} \e = -\e \label{projection}\eea The first
one is the usual D4-brane supersymmetry condition even in flat
space, where as the second condition is a projection condition on
the spinors. By using (\ref{flat-D4}) and (\ref{projection}), all
the gravitino variations are satisfied leaving the following
equations to have a constant spinor as a solution. \bea \p_0 \e =
0, \>\>\> \p_{\alpha} \e = 0,\>\>\> (\alpha = 3,\cdots,9), \>\>\>
\p_i \e + \frac{\mu}{2}J_{ij}\Gamma^{\hat j \hat 9} \e = 0, \>\>\>
(i =1, 2) \ . \eea Hence the D4-brane in the $n=1$ G${\rm \ddot
o}$del model preserves 1/4 of the total spacetime supersymmetry.
Similarly one can analyze the spacetime supersymmetry of the
D2-brane presented in (\ref{godel-D2}) by solving the dilatino and
gravitino variations.

\sect{Conclusions} We have presented in this paper a class of
G${\rm{\ddot o}}$del universe backgrounds from non-standard
intersecting branes in supergravity. These supersymmetric
backgrounds are different from the already known ones due to the
presence of various constant NS-NS and R-R field strengths. We
have further presented the supergravity solutions of D-branes in
type IIA theory in some G${\rm{\ddot o}}$del universes which are
obtained from the corresponding PP-wave backgrounds. The
supersymmetry properties of these branes are analyzed in detail by
solving the dilatino and gravitino variations explicitly. The
worldsheet construction of these branes can be carried out by
following \cite{Skenderis:2002wx}\cite{Harmark:2003ud} and looking
at the mixed boundary conditions properly. It will be interesting
to completely classify all the supersymmetric branes in
G${\rm{\ddot o}}$del universes of various kind. \vskip .1in
\noindent {\bf Acknowledgements:} We would like to thank S. F.
Hassan and Aalok Misra for various useful discussions. KLP would
like to thank the Abdus Salam I.C.T.P, Trieste for hospitality
under Associate Scheme, where a part of this work was completed.

\end{document}